\documentclass[12pt,a4paper]{article}
\usepackage{amsmath,amssymb,amsthm,mathptmx}
\usepackage[mathcal]{eucal}
\usepackage{url}
\usepackage{fullpage}

\newtheorem{theorem}{Theorem}

\author{Andrey Yu. Rumyantsev}
\title{Infinite computable version of Lovasz Local Lemma.
\thanks{Supported by RFBR 0901-00709a and NAFIT ANR-08-EMER-008 grants.}}

\date{}

\def\eps{\varepsilon}

\def\vbl{\mathrm{vbl}}

\begin{document}
\maketitle

\begin{abstract}
Lov\'asz Local Lemma (LLL) is a probabilistic tool that
allows us to prove the existence
of combinatorial objects in the cases when standard probabilistic
argument does not work (there are many partly independent
conditions).

LLL can be also used to prove the consistency
of an infinite set of conditions, using standard compactness argument
(if an infinite set of conditions is inconsistent, then some finite part
of it is inconsistent, too, which contradicts LLL).
In this way we show that objects satisfying all the conditions do exist
(though the probability of this event equals~$0$).
However, if we are interested in finding a \emph{computable}
 solution that
satisfies all the constraints, compactness arguments do not work
anymore.

Moser and Tardos~\cite{moser} recently gave
a nice constructive proof of LLL.
Lance Fortnow asked whether one can apply
Moser--Tardos technique
to prove the existence of a computable solution. We show that
this is indeed possible (under almost the same conditions as used
in the non-constructive version).
\end{abstract}

\section{Computable LLL: the statement.}

Let $\mathcal{P}$ be a sequence of mutually
independent random variables; each of them has a finite range.
(In the simplest case $P_i$ are independent random bits.)

We consider some family $\mathcal A$ of \emph{forbidden events};
each of them depends on a finite
set of variables, denoted $\vbl(A)$ (for event $A$).
Informally speaking, the classical LLL together with the
compactness argument guarantee that if
the events are of small probability and each of them is mostly independent
with the others, there exists an evaluation for all variables that avoids all the
forbidden events.

To make the statement exact, we need to introduce some terminology and
notation. Two events $A$ and $B$ are \emph{disjoint} if they do not share variables,
i.e., if $\vbl(A)\cap\vbl(B)=\varnothing$. For every $A\in\mathcal{A}$ let
$\Gamma(A)$ be the open (punctured) neighborhood of $A$, i.e., the set
of all events $E\in \mathcal{A}$ that share variables (are not disjoint)
with $A$, except $A$ itself.

\begin{theorem}[Infinite version of LLL]
     \label{infinite}
Suppose that for every event $A\in\mathcal{A}$ a rational number $x(A)\in (0,1)$
is fixed such that
    $$
\Pr [A]\le x(A)\prod_{E\in\Gamma(A)}(1-x(E)),
    $$
for all $A\in\mathcal{A}$. Then there exists an evaluation of variables that
avoids all $A\in\mathcal{A}$.
\end{theorem}

This is just a combination of finite LLL and compactness argument. Indeed,
each event from $\mathcal{A}$ is open the the product topology; if the claim
is false, these events cover the entire (compact) product space, so there exists
a finite subset of events that covers the entire space, which contradicts the
finite LLL.

Our goal is to make this theorem effective. For that we assume that we have
a countable sequence of variables $\mathcal{P}=P_0,P_1,\ldots$, the range of
$P_i$ is $\{0,1,\ldots,n_i-1\}$, and $n_i$ and the probability
distribution of $P_i$ are computable given $i$. Then we
consider a sequence of events, $\mathcal{A}=\{A_0,A_1,\ldots\}$. We assume
that these events are effectively presented, i.e., for a given $j$ one can compute
the list of all the variables from $\vbl(A_j)$ and the event itself (i.e., the list
of evaluations that belong to that event).  Moreover, we
assume that for each variable $P_i$ only finitely many events involve this
variable, and the list of those variables can be computed given $i$.

\begin{theorem}[Computable version of LLL]
\label{computablelll}
Suppose there is a rational constant $\eps\in(0,1)$ and a
computable assignment
of rational numbers $x:\mathcal{A}\to(0,1)$ such that
     $$
\Pr [A]\le (1-\eps)x(A)\prod_{E\in\Gamma(A)}(1-x(E)),
    $$
for all $A\in\mathcal{A}$. Then there exists a computable evaluation of variables that
avoids all $A\in\mathcal{A}$.
\end{theorem}

Note that the computability restrictions look quite naturally and that we only need
to make the upper bounds for probability just a bit stronger multiplying all the
bounds by some fixed constant $1-\eps$. (It should not be a problem for typical
applications of LLL; usualy this stronger bound on $\Pr[A]$ can be easily
established.)

\section{The proof}

To explain the proof, we recall first how Moser and Tardos prove the finite LLL.
(We do not repeat the argument here and assume that the reader is familiar with~
\cite{moser}: some estimates from this paper are needed and we assume that the
reader knows their proofs from~\cite{moser}.)

The probabilistic algorithm used in~\cite{moser} for the finite case, is quite natural:
it starts by assigning random values to all variables. Then, while there are some
non-satisfied conditions (=some bad events happen), the algorithm takes one of these
events
and resamples all the variables that appear in this
event (assigning fresh random values to them).

There is some freedom in this algorithm: the event for resampling can be chosen in
an arbitrary (deterministic or probabilistic) way.

\smallskip
We modify this algorithm for the case of infinitely many variables and events.
First we construct a probabilistic algorithm that with probability $1$ generates a
satisfying assignment in the limit (with predictable convergence, see below the
exact definitions). Then we use the existence of such an algorithm to show that
there is a computable assignment that satisfies all the conditions.

\smallskip
The probabilistic algorithm is a natural modification of Moser--Tardos algorithm.
We introduce some priority on conditions. For each condition we look at
the variables it involves, and take the variable with maximal index.
Then we reorder all the conditions in such a way
that
	$$
\max\vbl(A_0)\le \max\vbl(A_1)\le \max\vbl(A_2)\le\ldots
	$$
(Recall that each variable is used only in finitely many conditions, so we can make
the rearrangement in a computable way. This rearrangement is not unique.)

Then the algorithm works exactly as before, and we choose the first
violated condition (in this new ordering).

Remark: for some $n$ consider all the conditions that depend on variables
$P_0,P_1,\ldots,P_n$ only. These conditions form a prefix in our ordering.
Therefore, while not all of them are satisfied, we will not consider the other
conditions, so our infinite algorithm will behave (up to some point) like
a Moser--Tardos finite algorithm. They give a bound $x(A)/(1-x(A))$ for
an average number of resamples for condition $A$, so the expected total number
of resamples for this finite algorithm is finite. We come to the following
conclusion:

\textbf{Lemma 1}. \emph{With probability $1$ our algorithm will at some point
satisfy all the conditions depending on $P_0,\ldots,P_n$}.

\smallskip
Therefore, with probability $1$ the actions of the infinite probabilistic algorithm
can be split into stages: at $i$th stage we resample conditions that depend on
$P_0,\ldots,P_i$ only until all of them are satisfied. Let $p_0^i,\ldots,p_i^i$
be the values of the variables $P_0,\ldots,P_n$ at the end of the $i$th stage,
i.e., at the first moment when all the conditions depending only on $P_0,\ldots,P_i$
are satisfied.

These $p_i^j$ are random variables defined with probability $1$ (due to Lemma 1).
The values $p^i_0,\ldots,p^i_0$ form a satisfying assignment for all the conditions
that depend only on them. However, these values are not ``final'': when we start to
work with other variables, this may lead to changes in the previous variables.
So, e.g., $p_i^{j+1}$ can differ from $p_i^j$.

The compactness argument (that proves the existence of a satisfying assignment
for all condition) then takes the limit point of these assignments. This is not
enough for us, we need the following

\textbf{Lemma 2}. \emph{For every $i$ with probability $1$ the sequence
$$p_i^i, p_i^{i+1}, p_i^{i+2},\ldots$$
stabilizes.}

Moreover, for every variable with probability $1$ there exists some
moment in our algorithm such that after this moment it will never be changed.
(This is formally even a stronger statement since a variable can change during some
stage but return to its previous value at the end of the stage.)

\emph{Proof} of Lemma 2.
It is
enough to show that for every $i$ and sufficiently large $j$ the probability of them
event ``value of $P_i$ is changed after stage $j$'' is small.
To show this, we need to refer to the details of Moser--Tardos argument.
Consider all the events that involve the variable $P_i$. Then consider
all the neighbors of these events, all neighbors of their neighbors, etc.
($m$ times for some large $m$). Let $j$ be the maximal variable that
appears in all these events (up to distance $m$).

We claim that \emph{for every event $A$ that involves $P_i$, the probability of
being resampled after stage $j$ does not exceed $(1-\varepsilon)^m$}. Indeed,
consider such a resample and its tree (constructed as in \cite{moser}). This
tree should contain some event that involves variable with index greater than $i$
(since a new resample became necessary after all variables up to $P_j$ have
satisfactory values). The choice of $j$ guarantees then that the size of the
tree is at least $m$, and the sum of probabilities of all those trees to appear
during the algorithm is bounded by $(1-\varepsilon)^m x(A)/(1-x(A))$. By a
suitable choice of $m$ we can make this probability as small as we wish. Lemma
2 is proven.

Note that at this stage we have shown the existence of an evaluation (=assignment)
that satisfies all the conditions, since such an assignment is produced by
our algorithm with probability $1$. To show that there exists a \emph{computable}
assignment, we need some additional work.

\textbf{Lemma 3}. The convergence in Lemma 2 has predictable speed: for every $i$
and for every $\varepsilon$ one can compute some $N(i,\varepsilon)$ such that the
probability of the event ``value of $P_i$ will change after $N(i,\varepsilon)$ steps of
the algorithm'' is less than $\varepsilon$.

\textbf{Proof} of Lemma 3. The estimate in the proof of Lemma 2 gives some bound
in terms of the number of stages. At the same time we know the bounds for the
expected length of each stage, and can use Chebyshev inequality. Lemma 3 is proven.

Lemma 2 allows us to define an almost everywhere defined mapping that maps the
Cantor space $\Omega=\{0,1\}^\mathbb{N}$ into evaluations and maps the sequence of
random bits used by our algorithm to the sequence
$(p_0^\infty,p_1^\infty,\ldots)$ of limit values of the variables.

Lemma 3 guaranteed that the output distribution of this mapping (the image of
the uniform distribution on sequences of random bits) is computable. This means
that the probability of the event $p_0^\infty=a_0,\ldots,p_s^\infty=a_s$ can be
effectively computed (with any given precision) given $s$ and $a_0,\ldots,a_s$.
Indeed, due to Lemma 3 we know how many steps of the algorithm are needed to
get the output value with given certainty level, and can simulate our algorithm
for this number of steps. (Here we use the computability assumptions.)

This computable output distribution is concentrated on the set of satisfying
assignments. It remain to use the following simple remark.

\textbf{Lemma 4}. \emph{If a computable probability distribution is concentrated
on some closed set \textup{(}i.e. the measure of its complement is zero\textup{)},
then this set contains a computable element}.

Proof. Computing this distribution, we can choose sequentially the values
$a_0,a_1,a_2,\ldots$ in such a way that the measure of the event $p_0=a_0$, \ldots,
$p_k=a_k$ (according to the distribution) is positive for every $k$. The sequence
$a_0,a_1,a_2,\ldots$ is computable; if it does not belongs to the closed
set, then finitely many $a_0,\ldots,a_k$ ensure this, and this contradicts
the assumption (the probability should remain positive). Lemma 4 is proved,
and this finishes the proof of Theorem~\ref{computablelll}.

\section{Infinite CNFs}

A standard illustration for LLL is the following result:
\emph{a CNF where all clauses contain $m$ different variables and each clause
has at most $2^{m-2}$ neighbors, is always satisfiable}.

Here neighbors are clauses that have common variables.

Indeed, we let $x(A)=2^{-m+2}$ and note that
  $$
2^{-m} \le 2^{-m+2} [(1-2^{-m+2})^{2^{m-2}}],
  $$
since the expression in square brackets is approximately $1/e>1/2^2$.

This was about finite CNFs;
now we may consider effective infinite CNF with countably many variables and
clauses (numbered by natural numbers); we assume that for given $i$ we
can compute the list of clauses where $i$th variable appears, and for given
$j$ we can compute $j$th clause.

\begin{theorem}
    For every effective infinite CNF where each clause contains $m$ different variables
    and every clause has at most $2^{m-2}$ neighbors, one can find a computable
    assignment that satisfies it.
\end{theorem}

Indeed, the same choice of $x(A)$ works, if we choose $\varepsilon$ small enough
(say, $\varepsilon=0.1$).

Similar argument can be applied in the case where there are clauses
of different sizes. The condition now is as follows: for every variable there
are at most $2^{\alpha n}$ clauses of size that involve this variable,
where $\alpha\in(0,1)$ is some constant. Note that here we do not assume
that every variable appears in finitely many clauses, so the notion of
effective infinite CNF should be extended. Instead, we assume that for each
$i$ and for each $n$ one can compute the list of clauses of size $n$ that
include $x_i$.

\begin{theorem}\label{variable-cnf}
    For every $\alpha\in(0,1)$ there exists some $N$ such that every
    effective infinite CNF where each variable appears in at most $2^{\alpha n}$
    clauses of size $n$ \textup(for every $n$\textup) and all clauses have
    size at least $N$, has a
    computable satisfying assignment.
\end{theorem}

\emph{Proof}. Let us consider first a special case when each variable appears
only in finitely many clauses. Then we are in the situation covered by
Theorem~\ref{computablelll}, and we need only to choose the values of $x(A)$.
These value will depend on the size of the clause $A$: let us choose
  $$
x(A)=2^{-\beta k}
  $$
for clauses of size $k$, where $\beta$ is some constant. In fact, any constant
between $\alpha$ and $1$ will work, so we can use, e.g., $\beta=(1+\alpha)/2$.
So we need to check (for clauses of some size $k$) that
  $$
2^{-k} \le 2^{-\beta k} \prod_{B\in \Gamma(A)} (1-2^{-\beta\#B})
  $$
Note that for every of $k$ variables in $A$ there are at most $2^{\alpha m}$
clauses of size $m$ that involve it. So together there are at most $k2^{\alpha m}$
neighbors of size $m$. So it is enough to show that
  $$
2^{-k} \le 2^{-\beta k} \prod_{m\ge N} (1-2^{-\beta m})^{k2^{\alpha m}}
  $$
Using that $(1-h)^s \ge 1-hs$ and taking $k$th roots, we see that it is enough
to show that
  $$
2^{-1} \le 2^{-\beta} (1-\sum_{m\ge N}2^{\alpha m} 2^{-\beta m})
  $$
Since the series $\sum 2^{(\alpha-\beta)m}$ is converging, this is
guaranteed for large $N$.

So we have proven Theorem~\ref{variable-cnf} for the special case when
each variable appear only in finitely many clauses (and we can compute
the list of those clauses).

The general case is easily reducible to this special one. Indeed, fix
some $\delta>0$ and delete from each clause $\delta$-fraction of its variables
with minimal indices. The CNF becomes only harder to satisfy. But if
$\delta$ is small enough, the conditions of the theorem (the number
of clauses with $m$ variables containing a given variable is bounded
by $2^{\alpha n}$ are still true for some $\alpha'\in (\alpha,1)$.
And in this modified CNF each variable appears only in clauses of
limited size (it is deleted from all large enough clauses).

Theorem~\ref{variable-cnf} is proven.

Let us note some immediate corollaries. Assume that $F$ is a
set of binary strings that contains at most $2^{\alpha n}$ strings
of size $n$. Then one can use LLL to prove the existence of
an infinite (or bi-infinite) sequence $\omega$ and a number $N$ such
that $\omega$ does not have substrings in $F$ of length greater than $N$.
There are several proofs of this statement; one may use LLL or Kolmogorov
complexity, see~\cite{rumyantsev-1,rumyantsev-2}.

Joseph Miller noted that his proof (given in~\cite{miller-two-notes})
can be used to show that for a decidable $F$ (with this property) one
can find a computable $\omega$ that avoids long substrings in $F$.
Konstantin Makarychev extended this argument to bi-infinite strings
(personal communication). Now we get it as an immediate corollary
of Theorem~\ref{variable-cnf}: places in the sequence correspond
to variables, each forbidden string gives a family of clauses
(one per position), and there is at most $n2^{\alpha n}$ clauses
of size $n$ that involve given position (and this number is bounded by $2^{\alpha'n}$
for slightly bigger $\alpha'$ and large enough $n$).

Moreover, we can do the same for 2-dimensional case: having a decidable
set $F$ of rectangular patterns that contains at most $2^{\alpha n}$ different
patterns of size (=area) $n$, one can find a number $N$ and computable 2D configuration
(a mapping $\mathbb{Z}^2\to\{0,1\}$) that does not contain patterns from $F$ of
size $N$ or more. (The author does not know how to get this result directly,
not using Moser--Tardos algorithm.)

Author is grateful to Lance Fortnow who suggested to apply Moser--Tardos
technique to the infinite computable version of LLL.

\end{document}